\documentclass[12pt,english]{article}\usepackage[]{graphicx}\usepackage{xcolor}
\makeatletter
\def\maxwidth{ %
  \ifdim\Gin@nat@width>\linewidth
    \linewidth
  \else
    \Gin@nat@width
  \fi
}
\makeatother

\definecolor{fgcolor}{rgb}{0.251, 0.251, 0.282}

\usepackage{framed}
\makeatletter
 {\par\unskip\endMakeFramed%
 \at@end@of@kframe}
\makeatother

\definecolor{shadecolor}{rgb}{.97, .97, .97}
\definecolor{messagecolor}{rgb}{0, 0, 0}
\definecolor{warningcolor}{rgb}{1, 0, 1}
\definecolor{errorcolor}{rgb}{1, 0, 0}
\newenvironment{knitrout}{}{} 

\usepackage{alltt}
\usepackage[a4paper]{geometry}
\usepackage{natbib}
\usepackage{algorithm}
\usepackage{graphicx}
\usepackage{alphalph}

\usepackage[utf8]{inputenc}
\usepackage[T1]{fontenc}
\usepackage{tikz}
\usepackage{float}
\usepackage[colorlinks = true,linkcolor = blue, citecolor = blue]{hyperref}
\usepackage{booktabs}
\usepackage{amssymb,amsmath}
\usepackage{bbm}
\def\aaa{a}
\def\aaag{\textbf{a}}
\def\R{\mathbb{R}}
\def\alphag{\boldsymbol{\alpha}} 
\def\pig{\boldsymbol{\pi}} 
\def\psig{\boldsymbol{\psi}} 
\def\Hb{{\bf H}}
\def\ab{{\bf a}}
\def\hb{{\bf h}}
\def\ib{{\bf i}}
\def\zb{{\bf z}}
\def\ub{{\bf u}}
\def\xb{{\bf x}}
\def\Ab{{\bf A}}
\def\Bb{{\bf B}}
\def\Xb{{\bf X}}
\def\1b{{\bf 1}}

\newtheorem{exmpl}{Example}[section]
\newcommand{\myalphafoot}
{
\renewcommand{\thefootnote}{\alph{footnote}}
}

\title{Enhanced Cube Implementation \\ For Highly Stratified Population}
\myalphafoot
\author{\myalphafoot Rapha\"el Jauslin\footnotemark[1]~, Esther Eustache\footnotemark[1]~ and Yves Till\'e\footnotemark[1]}
\date{}
\footnotetext[1]{Institute of statistics, University of Neuch\^atel, Bellevaux 51, 2000 Neuch\^atel, Switzerland (E-mail: raphael.jauslin@unine.ch)}

\IfFileExists{upquote.sty}{\usepackage{upquote}}{}
\begin{document}

\maketitle

\begin{abstract}
A balanced sampling design should always be the adopted strategies if auxiliary information is available. Besides, integrating a stratified structure of the population in the sampling process can considerably reduce the variance of the estimators. We propose here a new method to handle the selection of a balanced sample in a highly stratified population. The method improves substantially the commonly used sampling design and reduce the time-consuming problem that could arise if inclusion probabilities within strata do not sum to an integer.  
\end{abstract}



\newpage


\section{Introduction}

In survey statistics, balanced sampling is a particularly efficient method when values of auxiliary variables are available for all units in the population. The idea is to select the sample so that the totals of the Horvitz-Thompson estimators of some auxiliary variables equal the population totals. There are different methods for selecting a balanced sample. \cite{dev:til:04a} have proposed the cube method which successively transforms the vector of inclusion probabilities into a sample. The method has been improved by \cite{cha:til:06} by reducing the computation time.

In many areas, it is very useful to use stratified sampling designs. As already indicated by \cite{ney:34}, the variance of the Horvitz-Thompson estimator can be reduced by constructing strata such that the variables are homogeneous within the strata.
Besides, \cite{cha:09} proposed a specific algorithm to obtain balanced samples in the strata of a population. However, this method becomes cumbersome when the number of strata is large.

A highly stratified population is very common in survey sampling. For example, it may be necessary to select individuals from a population while requiring that at most only one individual from each household in a population is taken.  Each household is then a stratum. In spatial statistics, one can also construct small strata of neighboring units to obtain well-spread samples. Highly stratified sampling is also necessary for some donor imputation methods: the objective is to select a respondent for each non-respondent to impute its values. Each non-respondent then defines a stratum from which a respondent is to be selected \citep{hasl:till:2016}.

The balanced and stratified sampling method of \citet{cha:09} has been improved by \cite{hasl:till:2014} to partially resolve the disadvantage of the time required to process a highly stratified population. When the sum of the inclusion probabilities in the strata is not an integer, the computation time can become problematic. This problem arises, for example, when the objective is to select less than one individual per household. Neither of the two methods already proposed solves the computational time problem in these situations.

In this paper, we propose a new method to obtain a stratified balanced sample. This new method is particularly interesting when the population is highly stratified and the inclusion probabilities do not sum to an integer within the strata. We refer readers to \citet{till:2020} and \citet{hank:mohr:newm:2020} to have more information on the general settings on stratified balanced sampling design.

The document is organized as follows. The section~\ref{sec:not} gives the basic notations and settings. Section \ref{sec:stratbal} present the problem of selecting a balanced sample. In the section~\ref{sec:cube}, we review the cube method and how it is used to select a balanced sample. In section~\ref{sec:high}, we discuss the issue of the highly stratified population and review the methods used to select a sample in this case. In the section~\ref{sec:strat}, we present the new method and the section~\ref{sec:var} is devoted to variance estimation. In the section~\ref{sec:simu}, we give the simulation results of the different algorithms on an artificial dataset while the section~\ref{sec:diss} gives a conclusion on the new method.


\section{Basic sampling notations}\label{sec:not}

Consider a finite population $U$ of size $N$ whose units can be defined by labels $k\in\{1,2,\dots,N\}$.
Let define a variable of interest $y$. Suppose that we are trying to estimate the following unknown total:
\begin{equation}\label{eq:Y}
Y = \sum_{k \in U} y_k.
\end{equation}

A sampling design is defined by the probability $p(s)$ of selecting each possible subset $s \subset U$ such that
$
\sum_{ s\subset U}p(s) = 1.
$
Consider a vector $\aaag = (\aaa_1,\dots,\aaa_N)^\top$ that maps elements of a subset $s$ to an $N$ vector of 0s and 1s such that: 
$$
\aaa_k =
\left\{\begin{array}{lll} 1 & \text{ if } k\in s, \\ 0 & \text{ otherwise} , \end{array} \right.
$$
for $k \in U$. For each unit of the population, the inclusion probability $\pi_k$, with $0\leq\pi_k\leq 1$, is defined as the probability of selecting $k$ into a sample $s$:
\begin{equation*}\label{eq:pik}
 \pi_k = \textrm{P}(k \in s) = \textrm{E}(\aaa_k) =  \sum_{s\subset U | k \in s} p(s).
\end{equation*}

Let $\pig=(\pi_1,\dots,\pi_N)^\top$ be the vector of all the inclusion probabilities. Let also $\pi_{k\ell}$ be the probability of selecting units $k$ and $\ell$ together in the sample, with $\pi_{kk} = \pi_k$.
Assuming that $\pi_k > 0$ for all $k\in U$, the total \eqref{eq:Y} can be estimated using the classical unbiased Horvitz-Thompson estimator defined by
\begin{equation}\label{eq:HT}
\widehat{Y} = \sum_{k\in U} \frac{y_k a_k}{\pi_k}.
\end{equation}


\section{Stratified balanced sampling} \label{sec:stratbal}

Usually, some auxiliary information are available for each unit $k\in U$ in a vector $\xb_k = (x_{k1},\dots,x_{kq})^\top \in\mathbb{R}^q$, with $q \in \mathbb{N^*}$.
A sampling design is said to be balanced on the $q$ auxiliary variables if and only if it satisfies the following balancing equation:
\begin{equation*}\label{eq:balance}
  \widehat{\Xb} = \sum_{k\in s} \frac{\xb_k}{\pi_k}  =  \sum_{k\in U} \frac{\xb_k a_k}{\pi_k}= \sum_{k\in U} \xb_k = \Xb.
\end{equation*}
Sometimes, selecting a sample that satisfies exactly the constraints is not possible due to the rounding problem.

In many applications, inclusion probabilities are such that the selected sample has a fixed size.
In order to obtain a sampling design with fixed sample size, a linear combination of the auxiliary variables must be proportional or equal to the vector of inclusion probabilities, i.e. there exists $\psig \in \R^q$ such that $\psig^\top \xb_{k} = \pi_{k}$, for all $k \in  U$. Indeed, this gives
$$ \sum_{k\in s} \frac{\psig^\top \xb_{k}}{\pi_{k}} = \sum_{k \in s}\frac{\pi_{k}}{\pi_{k}} = n. $$ 
The size of the sample will be fixed only if $n$ is an integer.
If it is not the case, the sample size will be equal to the higher or lower integer to $n$. 

More generally, the problem of selecting a balanced sample is written as the following linear system :
\begin{equation}
\label{eq:bal}
\left\{\begin{array}{lll}
\displaystyle \Ab^\top\ab = \Ab^\top\pig,\\
\ab \in\{0,1\}^N,
\end{array}\right.
\end{equation}
where $\Ab= \left( \xb_1 / \pi_1,\dots, \xb_N / \pi_N \right)^\top$.
The aim consists then of obtaining a sample $\ab$ that satisfies (or approximately satisfies) the constraints.

Suppose that the population $U$ is divided into $H$ strata $U_1,\dots,U_H$, with respective sizes of $N_1,\dots,N_H$. The strata form a partition and respect the following properties:

$$U = \bigcup_{h = 1}^H U_h,~ N_h > 0,~ U_h\cap U_\ell = \emptyset, \text{ for all } h,\ell \in \{1,\dots,H\}$$
Then, this implies that $\sum_{h = 1}^H$ $N_h = N$.
The inclusion probabilities sum to a value $n_h$ in each stratum $h$, i.e. $n_h = \sum_{k \in U_h} \pi_k$.
Let $\hb = ( h_1,$  $\dots,$ $h_N)^\top$ be a categorical vector that specifies the stratum to which each unit belongs.
For example, $h_k = \ell$ means that unit $k$ belongs to strata $U_\ell$, with $k \in U$ and $\ell \in \{1,\dots,H\}$.
Another way for expressing the stratum of each unit is to use the disjunctive form.
Let $\Hb$ be the disjunctive matrix of the corresponding vector $\hb$ of size $N\times H$, such that:
$$\displaystyle \Hb = \big( \1b (U_1), \dots, \1b (U_H) \big), $$
where $\1b (U_h) \in \mathbb{R}^N$ is an column vector such that its $k$th element is equal to 1 if the unit $k$ belongs to the stratum $U_h$ and 0 otherwise.

Obtaining a balanced sample in a stratified population is equivalent to adding stratification constraints to the previous linear system~\eqref{eq:bal}.
These constraints are contained in the matrix $\Hb$, so the modification of the linear problem gives:
\begin{equation}
\label{eq:bal2}
\left\{\begin{array}{lll}
\displaystyle \left( \Hb \ \Ab \right)^\top \ab = \left( \Hb \ \Ab \right)^\top \pig,\\
\ab \in\{0,1\}^N.
\end{array}\right.
\end{equation}
The number of constraints in the linear problem is then $\left( q+H \right)$.
In the next section, a method to select a balanced sample is presented.


\section{Cube Method}\label{sec:cube}

\cite{dev:til:04a} developed the cube method that selects a balanced sample respecting the inclusion probabilities. The method can deal with equal or unequal inclusion probabilities.
The algorithm is separated into two phases.

\begin{itemize}
\item The first phase is called the flight phase.
It modifies recursively and randomly the vector of inclusion probabilities $\pig$ into a sample by respecting exactly the balancing constraints of the problem.
The subspace induced by the linear system \eqref{eq:bal} could be rewritten using the following notation:
$$\mathcal{A} = \left\{ \ab \in \R^N | \Ab^\top \ab = \Ab^\top \pig \right\} = \pig + \text{Null}(\Ab^\top),$$
where $\text{Null}(\Ab^\top) = \left\{\ub \in \R^N | \Ab^\top\ub = 0 \right\}$.
The idea is then to use a vector $\ub$ of the null space of $\Ab^\top$ in order to update randomly the vector $\pig$. 
The whole procedure of the update can be found in \cite{dev:til:04a}.
At each step, at least one component is set to 0 or 1. Matrix $\Ab$ is updated with the new inclusion probabilities.
This step is repeated until the null space of $\Ab^\top$ is empty.
At the end of the flight phase, the final updated vector of $\pig$ contains at most $q$ elements that are still not equal to 0 or 1.
\item The second phase is called the landing phase.
This phase allows to obtain the sample $\ab$ that respects as much as possible the balancing constraints.
There are two different ways to achieve it, by relaxing the $q$ constraints one by one, or by linear programming.
\end{itemize}
In the flight phase, the major computational cost comes from the research of a vector in the null space of $\Ab^\top$.
\cite{cha:til:06} have improved this time-consuming inconvenience using a sub-matrix of $\Ab$ rather than the entire matrix.
The idea is to consider a submatrix that has one more row than the number of columns to ensure to have at least one vector in its null space.
This submatrix, denoted by $\Bb$, has then a size of $(q+1)\times q$, with respect to $q<N$ and $\mbox{Rank}(\Bb) \leq q$.

The interest of using this submatrix comes from the following result: a vector $\ub$ of $\text{Null}(\Bb^\top)$ completed by $\left( N-(q+1) \right) $ zeros is a vector of $\text{Null}(\Ab^\top)$.
With this idea, all the computations can be done using only a submatrix $\Bb$.
Usually, $N$ is much greater than $q$, the size of $\Bb$ is then much smaller than $\Ab$. This implies obviously an important gain of computational time.
The method proposed in this paper uses the same idea.
In the next section, the particular case of highly stratified sampling is considered.

\section{Highly stratified population}\label{sec:high}
It is always preferable to consider a stratified population in order to estimate the total~(\ref{eq:Y}).
Indeed, the variance of the Horvitz-Thompson estimator~(\ref{eq:HT}) can be considerably reduced compared to the non-stratified estimator~(\ref{eq:Y}).
However, when the population is highly stratified (i.e. $H$ is very large), the selection of a balanced sample with classical methods becomes difficult due to the too large number of constraints in $\Hb$.
In order to decrease the time-consuming problem, different approaches have already been proposed.

\citet{cha:09} has developed an algorithm to select a balanced sample in a highly stratified population.
Firstly, a flight phase is applied inside each stratum.
This allows modifying the inclusion probabilities such that these are as balanced as possible in each stratum.
Next, a flight phase is applied on the whole population.
Finally, a landing phase is carried out on units that are not still selected or rejected.
This procedure has the advantage to be simple to implement.
Its major deficiency is when the number of strata $H$ becomes too large, the procedure remains very slow and often cannot even be used.

\citet{hasl:till:2014} have proposed another method to deal with highly stratified population. As the previous method, it begins by applying the flight phase of the cube method to each stratum of the population.
Next, it carries out a flight phase on an union of strata by adding another stratum at each step.
By doing this, strata are managed one after the other and the inclusion probabilities of certain strata are set to 0 or 1 during this step.
The idea behind this procedure is to reduce the matrix $\Hb$ considered because some strata are removed from the matrix when all its units are selected or rejected.
At the end, a landing phase is applied.
However, if $n_h$ is not equal to an integer for a stratum $U_h$, this method also remains very time-consuming.
Indeed, some strata are never completely removed during the procedure and then the submatrix of $\Hb$ considered becomes too large.

The properties of the cube method imply that the inclusion probabilities are satisfied and that the sample is balanced on the auxiliary variables in these two methods.
However, they still have difficulty to deal with all the situations of highly stratified sampling.
In the next following section, a new method is presented in order to completely resolve these drawbacks.

%


\section{Proposed method}\label{sec:strat}

In the fast implementation of the cube method \citep{cha:til:06}, the main modification was to use a matrix smaller than $\Ab$ to update $\pig$. This allows to considerably decreasing the computational cost.
The idea of our method is similar but adapted to a stratified population: consider a matrix of constraints $\Bb$ smaller than $( \Hb \ \Ab )$ during the use of the cube method.

The submatrix matrix $\Bb$ must be found at each step of the flight phase of the cube method.
As explained in Section~\ref{sec:stratbal}, the number of balancing constraints depends on the number of strata $H$ when the population is stratified.
By considering a matrix $\Bb$ with fewer rows, or units, the corresponding vector of strata $\hb$ will be reduced.
This subvector of $\hb$ will contain fewer categories and then the corresponding matrix $\Hb$ will have fewer columns.
The number of constraints will therefore depend on the rows of $\Bb$.
This is why obtaining the matrix $\Bb$ with exactly one row more than its number of columns is not as easy as with an unstratified population.
Algorithm~\ref{algo:findB} shows how to find the number of rows to consider to obtain the smaller matrix $\Bb$ such that $\Bb$ has exactly one row more than its number of columns.

\begin{algorithm}
\caption{Find the sub-matrix $\Bb$ of $\left( \Hb \ \Ab \right)$}\label{algo:findB}
Let $q$ be the number of auxiliary variables of $\Ab$. Initialize $q^1$ by $q$. For $t = 1,2,3,\dots$ repeat the following steps:
\begin{enumerate}
\item Extract the first $q^t$ rows of the vector $\hb$ and denote it $\hb^t$.
\item Denote $H^t$ the number of different strata in $\hb^t$.
\item Update $q^{t+1} = q + H^t + 1$.
\end{enumerate}
while $q^{t+1} > q^t$.\\
Finally, $\Bb$ is defined as the $q^t$ first rows of the concatenated matrix $\left( \Hb^t \ \Ab^t \right)$, where $\Ab^t$ and $\Hb^t$ are the submatrix containing only its $q^t$ first rows.

\end{algorithm}


\begin{exmpl}
Suppose that $q = 2$ and that the categorical vector is equal to $\hb = (1,$ $1,$ $2,$ $2,$ $3,$ $3,$ $3,$ $4,$ $4)^\top$. We obtain
$$\begin{array}{llllll}
t = 1: & q^1 = 2,  & \hb^1 = (1,1)^\top,       & H^1 = 1 &\to& q^2 = 2+1+1 = 4, \\
t = 2: & q^2 = 4,  & \hb^2 = (1,1,2,2)^\top,       & H^2 = 2 &\to& q^3 = 2+2+1 = 5, \\
t = 3: & q^3 = 5,  & \hb^3 = (1,1,2,2,3)^\top,       & H^3 = 3 &\to& q^4 = 2+3+1 = 6, \\
t = 4: & q^4 = 6,  & \hb^4 = (1,1,2,2,3,3)^\top,       & H^4 = 3 &\to& q^5 = 2+3+1 = 6, \\
t = 5: & q^5 = q^4 
\end{array}$$
$\Bb$ contains then $q^4 = 6$ rows and 2 + 3 = 5 columns. So it is a matrix with only one more rows than the number of columns as desired.
\end{exmpl}

The matrix $\Bb$ is found after having computed its number of rows $q^t$ using Algorithm~\ref{algo:findB}. The first $q^t$ elements of $\hb$ composed the strata membership vector $\hb^t$. The disjunctive matrix $\Hb^t$ can then be found using $\hb^t$. The matrix $\Bb$ is equal to $\left( \Hb^t \ \Ab^t \right)$, with $\Ab^t$ the submatrix of $\Ab$ containing only its $q^t$ first rows. The same procedure proposed by \citet{cha:til:06} can then be applied.
If the population is highly stratified and the number of auxiliary variables is acceptable, our procedure can be very efficient.
Moreover, it handles inclusion probabilities that do not sum to an integer inside strata. Algorithm~\ref{algo:cube2} presents the whole method.

\begin{algorithm}
	\caption{}\label{algo:cube2}
	Consider $\pig$ the $N$ vector of inclusion probabilities such that $0< \pi_k <1$, for $k \in \{1,\dots,N\}$.

	\begin{itemize}
		\item[I.] Perform a flight phase on each stratum according to the inclusion probabilities $\pig$ and the balancing constraints in $\Ab^\top$.
		The vector $\pig$ is updated by $\pig^1$ such that some of its elements are set to 0 or 1.
		Compute the set of indices $\ib^1 \subset \{1,\dots,N\}$ containing the unit indices that have an inclusion probability still not equal to~0 or~1.
		\item[II.]
		Initialize $t$ by 1.
		Repeat step~1. to~6. until it is no more possible to find the matrix $\Bb$ or until the vector $\ub$ is null.
		\begin{enumerate}
			\item In $\Ab$, $\hb$ and $\pig$, consider only units with indices in $\ib^t$.
			\item Apply the Algorithm~\ref{algo:findB} to find the submatrix $\Bb$ of $(\Hb \ \Ab)$.
			\item 
			Compute $\ub$, a vector of the null space of $\Bb$ completed by 0s to obtain a vector with the same size as $\ib^t$.
			\item
			Compute $\lambda_1 > 0$ and $\lambda_2 > 0$, the two greater values such that
			$$
			\begin{array}{ccccc}
				0 &\leqslant & \pi_k^{t} + \lambda_1 u_k \leqslant 1\\
				0 &\leqslant & \pi_k^{t} - \lambda_2 u_k \leqslant 1\\
			\end{array}
			\text{, for all } k \in \ib^t.
			$$
			\item Update $\pig^t$ by:
			$$\pig^{t+1} = \left\{\begin{array}{cccc}
			\pig^{t} + \lambda_1 \ub & \text{ with probability } & \lambda_2/(\lambda_1 + \lambda_2),\\
			\pig^{t} - \lambda_2\ub & \text{ with probability } & \lambda_1/(\lambda_1 + \lambda_2).
			\end{array}\right.$$
			\item Update $t$ by $t+1$ and update $\ib^t$ the set of indices containing the unit that have an inclusion probability still not equal to~0 or~1.
		\end{enumerate}
		\item[III.]
		It could remain some units that are still not rejected or selected.
		Perform a landing phase by suppression of variables on the balancing variables $(\Hb \ \Ab)$ on the remaining indices $\ib^t$ .
	\end{itemize}
\end{algorithm}

\section{Variance estimation}\label{sec:var}

The variance can be approximated using the method proposed by \citet{dev:til:05}.
Let the vector
$$ \zb_k = (\Hb ~~ \Ab )_{k},$$
where $(\Hb ~~ \Ab )_{k}$ denote the $k$th row of the matrix $(\Hb ~~ \Ab )$.
The variance of the Horvitz-Thompson estimator of the total $\widehat{Y}$ can be approximated by
\begin{equation}\label{eq:varapp} \text{var}_{app}(\widehat{Y}) = \sum_{k\in U} c_k \left( \dfrac{y_k}{\pi_k} - \alphag^\top\zb_k\right)^2, \end{equation}
where
\begin{equation}\label{eq:varest}
c_k = \pi_k(1-\pi_k)\frac{N}{N-(H+q)}
\text{ and }
\alphag = \left( \sum_{\ell\in U}c_\ell \zb_\ell \zb_\ell^\top \right)^{-1} \sum_{\ell \in U} c_\ell \zb_\ell \dfrac{y_\ell^\top}{\pi_\ell}.
\end{equation}

There exists many different ways to express the quantity $c_k$ and then this leads to various approximations of the variance.
Value $c_k$ can in particular be approximated by $$\widetilde{c_k}=(1-\pi_k)\frac{n}{n-(H+q)}.$$
Equation \eqref{eq:varapp} can be estimated on a sample $s$ using $\widetilde{c_k}$ instead of $c_k$ and by replacing the sums on $U$ by sums on $s$ in Expressions~(\ref{eq:varapp}) and~(\ref{eq:varest}).


\section{Simulations}\label{sec:simu}

In this section, the performance of the method is evaluated on real data produced by the \citet{GEOSTAT}.
The dataset contains information on Swiss establishments.
We restrict the study to the Switzerland region called Espace Mittelland (a region of the second degree of the Nomenclature of Territorial Units for Statistics (NUTS) of Switzerland).
This region contains 5 cantons (a region of the third degree of the NUTS) and 675 municipalities.
For confidentiality reasons, the units considered are the hectares of land in which at least one establishment is located.
In order to be able to estimate the variance, only 3 hectares of land per municipalities are included in the study.
This implies that the dataset contains information from 2025 hectares including at least one establishment.

\begin{figure}[ht!]
 	\centering
 	\input{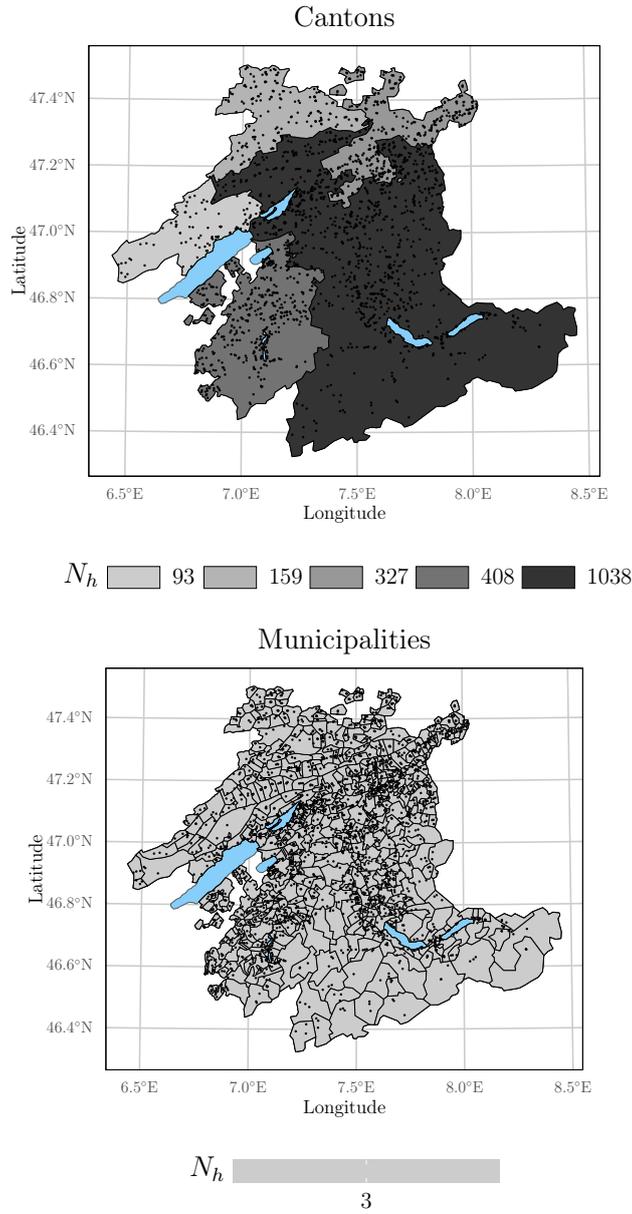}
 	\caption{Data extract from the Swiss establishement data base of the Swiss Federal Statistical Office \citet{GEOSTAT}. The data are restricted to the NUTS region 2. The left plot is showing the separation by Canton $H_c = 5$, the right one the separation by Municipality $H_m = 675$. The grey gradient scale give the number of units considered in each Canton. The data are selected such that each municipality contain 3 units.}
 	\label{fig:SwissPlot}
\end{figure}

We stratify the units in two different ways: by cantons and by municipalities. The number of strata is then respectively equal to $H_{c} = 5$ and $H_{m} = 675$. Figure \ref{fig:SwissPlot} shows the dataset with the two proposed stratification. The idea behind this procedure is to compare the execution time for a stratified population with a low number of strata versus a high one.
To compare the method, we will use balancing variables $\xb_j$ containing the number of women employed in a sector $j$, with $j = 1,2,3$.
Each sector represents a type of activity: sector $j=1$ involves the natural environment and agriculture, sector $j=2$ is manufacture and sector $j=3$ is related to services.
Table \ref{tab:time} shows the mean time execution of three methods for highly stratified sampling: the methods of \citet{hasl:till:2014} and \citet{cha:09}, detailed in Section~\ref{sec:high}, and the new one presented in this article.

Inside each stratum, the inclusion probabilities are equal.
For each stratification, we carry out two different sampling: one with inclusion probabilities that sum to an integer number within each stratum and one with a non-integer sum.
Then, we consider $n_h = 80$ and $n_h = 80.4$, $h \in \{ 1, \dots, H_c\}$, for the first stratification. For the second one, we take $n_h = 2$ and $n_h = 1.4$, $h \in \{ 1, \dots, H_m\}$.
We choose to deal also with non-integers $n_h$ with the aim to compare the impact of this situation on the mean sampling time.

Chauvet's method cannot be compared because its execution time is too long and should be avoided for highly stratified population. However, it remains very efficient if the number of strata is acceptable.
If inclusion probabilities in stratum sum to an integer, the Hasler's method performs very well. However, the execution time increases strongly when $n_h$ is not integer.
The proposed method has a very well behaved and the time is considerably reduced for a highly stratified population.

In order to compare the variance of the method with the others, we estimate the variance using some variables of interest $y_j$ that contains the total number of employee of the sector $j$, $j = 1, 2, 3$.
In Table~\ref{tab:time}, we compare the approximated variance \eqref{eq:varapp}, the estimated variance 
and the simulated variance computing using the equation:
\begin{equation}\label{eq:varsim}
	v_{sim} = \frac{1}{m}\sum_{s}\{\widehat{Y}(s) - Y \}^2,
\end{equation}
where $m$ is the number of simulations.

For each method, we vary the number of selected units within each stratum by taking $n_h$ equal to 2 for the stratification with $H_c$ strata and 80 for the stratification with $H_m$ strata. This implies sample of size respectively equal to 400 and 1350.
The variance estimator seems to be unbiased for the approximated variance. However, we see that the approximated variance and estimator are slightly biased to the $v_i$. This coming from the landing phase of each method.
We can conclude that the proposed method is comparable in terms of variance to other methods.

\begin{table}[!h]

\caption{\label{tab:time}Results of 1000 simulations on the Swiss establishments dataset. The population of size is equal to 2025. We compute the mean time execution in secondes of each sampling procedure. We vary the number of strata $H$ and the number of unit selected within each strata $n_h$.}
\centering
\resizebox{\linewidth}{!}{
\fontsize{7}{9}\selectfont
\begin{tabular}[t]{rrrr}
\toprule
\multicolumn{1}{c}{ } & \multicolumn{3}{c}{Algorithm} \\
\cmidrule(l{3pt}r{3pt}){2-4}
\multicolumn{1}{c}{ } & \multicolumn{1}{c}{Proposed method} & \multicolumn{1}{c}{Hasler's method} & \multicolumn{1}{c}{Chauvet's method} \\
\cmidrule(l{3pt}r{3pt}){2-2} \cmidrule(l{3pt}r{3pt}){3-3} \cmidrule(l{3pt}r{3pt}){4-4}
\addlinespace[1ex]
\multicolumn{4}{l}{Cantons ($H = 5$)}\\
\hspace{1em}$n_h = 80$ & 0.24 & 0.25 & 0.24\\
\hspace{1em}$n_h = 80.4$ & 0.24 & 0.24 & 0.24\\
\addlinespace[1ex]
\multicolumn{4}{l}{Municipalities ($H = 675$)}\\
\hspace{1em}$n_h = 2$ & 0.42 & 0.4 & 418.07\\
\hspace{1em}$n_h = 1.4$ & 0.53 & 400.55 & 701.77\\
\bottomrule
\end{tabular}}
\end{table}

\begin{knitrout}
\definecolor{shadecolor}{rgb}{0.973, 0.973, 0.973}\color{fgcolor}\begin{table}[!h]

\caption{\label{tab:v}Results of 1000 simulations on a population of size is equal to 2025. The number of strata is equal to 5 for Cantons and 675 for the Municipalities. For each variable of interest $y_j$, $j = 1,2,3$ and for each sampling methods, we compute the variance approximated by the simulations \eqref{eq:varsim} , approximated variance \eqref{eq:varapp} as well as the variance estimator \eqref{eq:varest}}
\centering
\resizebox{\linewidth}{!}{
\fontsize{9}{11}\selectfont
\begin{tabular}[t]{llllllllll}
\toprule
\multicolumn{1}{c}{ } & \multicolumn{9}{c}{Algorithm} \\
\cmidrule(l{3pt}r{3pt}){2-10}
\multicolumn{1}{c}{ } & \multicolumn{3}{c}{Proposed method} & \multicolumn{3}{c}{Hasle's method} & \multicolumn{3}{c}{Chauvet's method} \\
\cmidrule(l{3pt}r{3pt}){2-4} \cmidrule(l{3pt}r{3pt}){5-7} \cmidrule(l{3pt}r{3pt}){8-10}
 & $v_{sim}$ & $\widehat{\text{var}}(\widehat{Y})$ & $\text{var}_{app}(\widehat{Y})$ & $v_{sim}$ & $\widehat{\text{var}}(\widehat{Y})$ & $\text{var}_{app}(\widehat{Y})$ & $v_{sim}$ & $\widehat{\text{var}}(\widehat{Y})$ & $\text{var}_{app}(\widehat{Y})$\\
\midrule
\addlinespace[1ex]
\multicolumn{10}{l}{Cantons $(H = 5)$}\\
\hspace{1em}$y_1$ & 25788.227 & 24365.755 & 24735.596 & 24286.279 & 24147.542 & 24735.596 & 22913.046 & 24368.969 & 24735.596\\
\hspace{1em}$y_2$ & 136146.545 & 123257.656 & 131861.315 & 130340.065 & 121443.551 & 131861.315 & 134082.01 & 124254.374 & 131861.315\\
\hspace{1em}$y_3$ & 145849.824 & 129921.58 & 135199.514 & 148338.808 & 128108.505 & 135199.514 & 153747.141 & 129142.588 & 135199.514\\
\addlinespace[1ex]
\multicolumn{10}{l}{Municipalities $(H = 675)$}\\
\hspace{1em}$y_1$ & 1656.128 & 1726.496 & 1736.933 & 1815.448 & 1727.489 & 1736.933 & 1771.027 & 1731.304 & 1736.933\\
\hspace{1em}$y_2$ & 9177.576 & 9394.483 & 9435.185 & 9109.812 & 9427.651 & 9435.185 & 9887.571 & 9408.017 & 9435.185\\
\hspace{1em}$y_3$ & 9033.066 & 8990.18 & 8982.537 & 9589.876 & 8950.333 & 8982.537 & 10533.724 & 8937.991 & 8982.537\\
\bottomrule
\end{tabular}}
\end{table}

\end{knitrout}

\section{Conclusion}\label{sec:diss}

The stratified sampling procedure is a well-known and appropriate procedure to reduce the variance of the Horvitz-Thompson estimator.
In this paper, we propose a new method and implementation that provides an excellent executive time and flexibility that the existing methods did not allow.
In many surveys where the population is stratified, the sum of inclusion probabilities within each stratum is not an integer.
Other methods are not directly applicable in this case.
We have shown by mean of simulations that the variance of the estimator is not impacted by our method. All of these results indicate that our proposed algorithm is very efficient to select a sample in a stratified and highly stratified population.


\end{document}